# Early Stages of Flame Propagation in Tubes with No-slip Walls and the Mechanism of Tulip Flame Formation

Chengeng Qian [a], Cheng Wang [a] Michael A. Liberman [b]*

[a] State Key Laboratory of Explosion Science and Technology, Beijing Institute of Technology, Beijing, 100081, China
[b] Nordita, KTH Royal Institute of Technology and Stockholm University, Hannes Alfvéns väg 12, 114 21 Stockholm, Sweden

## Abstract

The early stages of flames propagation in tubes with no-slip walls and the inversion of the flame front from a convex shape directed towards unburned gas to a concave shape with a cusp directed to the burned gas, known as a tulip flame, was investigated for closed and half-open tubes by solving the fully compressible reactive Navier-Stokes equations with a one-step Arrhenius chemical model for the highly reactive hydrogen/air and slowly reacting methane/air mixtures. The development of the tulip flame in hydrogen/air obtained in simulations with a one-step Arrhenius model was compared with simulations using a detailed chemical model. It is shown that the inversion of the flame front and the onset of the tulip flame occurs due to rarefaction waves generated by the decelerating flame when its surface was reduced due to extinguishing of the rear parts of the flame skirt at the sidewalls. The rarefaction waves reduce the flow velocity ahead of the flame, creating an uneven velocity profile, which facilitate the flame front inversion. In the case of a flame propagating in a tube with both ends closed, the compression wave reflected from the tube end opposite to the ignition end, also contribute to the inversion of the flame front. In the case of fast flames and a relatively short tube with both closed ends, the rarefaction waves may invert the flow velocities in the unburned gas towards the propagating flame. The tulip flame formation mechanism is a purely gas-dynamic phenomenon not associated with flame instabilities or vortex motion, in agreement with the conclusion obtained in recent experimental studies [1].

* Corresponding author.
E-mail address: mliber@nordita.org; mliber@kth.se (M. Liberman).

# 1. Introduction

The dynamics of flames propagation in tubes is important for understanding combustion processes under confinement, such as explosions and safety issues as well as industrial and technological applications, e.g. combustion in gas turbines and internal combustion engines. The inversion of the shape of the flame front from a convex, directed towards the unburned gas to a concave shape with a cusp directed to the burnt gas, is well known, and the onset of tulip flame has been observed in many experiments and numerical simulations. From the beginning, the flame, which is ignited near the centerline at the closed end of the tube and propagates to the opposite closed or open end, takes a convex shape with a cusp directed towards the unburned gas. The flow in the unburned gas is caused by the thermal expansion of the high-temperature combustion products. The flow velocity in the unburned gas mixture is parallel to the tube walls, while the velocity profile is uniform in the bulk, and drops to zero on the side walls in the boundary layer. The flame surface increases due to the flame front is stretched along the boundary layer near the sidewalls, and the combustion velocity increases due to the increase in the flame surface. Since the rate of the flame surface increase caused by the flame surface stretching along the sidewalls is much less than the transit time of a gas particle across the flame front, $\tau_{tr} = L_f / U_f$, the internal structure of the flame can be considered as quasi-stationary. When the flame skirt touches the side walls of the tube, the rear part of the flame skirt is extinguished, which reduces the surface area of the flame front and, accordingly, the speed of the combustion wave decreases. The reduced expansion of the burnt gases and the deceleration of the flame cause rarefaction waves that propagate into the unburned gas, can be reflected from the opposite closed end towards the flame and penetrate into the burnt gas behind the flame. Soon after the flame surface area and the flame velocity decrease, the flame front begins to invert, forming a cusp pointing towards the burnt gas and lagging behind the leading edges of the flame front near the side walls.

## 2. A brief overview of previous investigations on tulip flame formation

The first experimental observation and photographic study of the "tulip" flame were published by Ellis with collaborators [2, 3, 4] almost a century ago using a newly invented at that time rotating shutter camera. Later, Salamandra et al. [5] called the inward pointed concave flame "tulip flame", and this term is commonly used to describe a specific cusped-shape flame that occurs during the flame propagation in tubes observed in various experimental and numerical studies, e.g. the deflagration-to detonation transition.

The formation of a tulip flame was observed when the flame was ignited near the closed end of the tube and propagates to the opposite closed or open end. In sufficiently long tubes with an aspect ratio $L/D$ greater than 20 and usually with both ends closed, the shape of the flame may at some point change again from a concave to a convex shape in the direction the unburned mixture, and later it may turn into a tulip flame again [6]. The most pronounced flame-inversion behavior occurs when a flame is ignited at the closed end of a tube and propagates to the opposite closed end.

Following the earliest publications by Ellis et al. [2-4], the formation of a tulip flame has been studied by many researchers. The first attempt to explain the mechanism of the tulip flame formation was undertaken by Markstein [7], who suggested that the inversion of the flame shape may be the result of the interaction of a curved flame front with a planar shock wave. Starke and Roth [8] hypothesized that the onset of the tulip flame in Markstein's experiments is associated with the Rayleigh-Taylor instabilities. However, the relatively high intensity of shock waves, such as in Markstein experiments, does not appear in the early stage of flame propagation, when the tulip shape is formed.

Of particular note is the comprehensive study of the tulip flame formation by Guénoche [6], who put forward the fruitful hypothesis that the flame generates rarefaction waves during the deceleration phase, which he considered as a key process in the tulip flame formation. Surprisingly, subsequent researchers did not pay due attention to the Guénoche's work [6].

A detailed information on the flow, pressure field and the flame evolution were obtained by Dunn-Rankin et al. in the series of experimental studies [9-12]. From his early experiments [10] Dunn-Rankin concluded that the flame front inversion is an initial distortion caused by the Darrieus- Landau instability, which grows, developing in the form of a cusp directed backward, thus forming a tulip shape. Later Dunn-Rankin [11, 12] interpreted the flame front inversion as associated with the recirculation of burned gases observed in schlieren images. It was hypothesized that the expansion of burned gases along the normal to the convex flame surface deflects the flow behind the flame towards the tube axis. Kerampran et al. [13] studied a flame in different combustible mixtures propagating from the closed to the open end of a tube and found that it was always accompanied by pressure oscillations. Hariharan and Wichman [14] carried out a detailed experimental study and two-dimensional numerical simulations of the flame propagating in a rectangular channel with both closed ends with an aspect ratio $L/D=6$. They concluded that the transition to the tulip shape is influenced by the "stagnation point" – when the flame front is almost planar and the flame speed is minimal. They emphasized that further inversion of the flame front and the development of a tulip shape are dictated by the presence of vortices in the channel.

Clanet and Searby [15] conducted extensive experimental research with a half-open tube. Their experiments showed that the time of the tulip flame formation mainly depends on the aspect ratio of the tube[1], $L/D$, which can be considered as an indication that the process is influenced by the structure of the boundary layer in the unburned gas ahead of the flame, and, possibly, by vortices into the boundary layer in burned combustion products behind the flame. At the same time, Clanet and Searby concluded that neither the boundary layer or acoustic effects are dominant in the formation of the tulip flame. While this conclusion may be correct for acoustic effects, the conclusion about the boundary layer is questionable because many experiments show a close relationship between the tulip flame formation and the upstream flow ahead of the flame. In addition, their suggestion that the Rayleigh-Taylor instability may be important in the tulip flame formation was not confirmed by subsequent experiments and simulations. It should be emphasized that the tulip flame has never been observed in narrow tubes, since the Poiseuille flow is very quickly established in the flow ahead of the flame, and the flame is accelerated all the time without a deceleration phase.

The quantitative data obtained by Dunn-Rankin [9, 10], Starke and Roth [8] and Jeung, Cho and Jeong [16] from velocity measurements during the tulip flame formation have stimulated a series of numerical simulations of the tulip phenomenon [17-19]. Various mechanisms have been proposed in an attempt to explain the inversion of the flame front leading to the tulip flame. One of the first hypotheses [6, 7, 20, 21] was based on the similarity of the tulip phenomenon and the inversion of the flame front as a result of a curved flame interaction with a shock wave. Various other hypotheses have been proposed, such as the effect of the Darrieus-Landau and Taylor instabilities [17, 18, 22-26] and the effect of a flame interaction with acoustic waves [25, 26]. It was believed [18, 22] that DL instability is always present in the flame front inversion, but this is incompatible with the fact that the characteristic time scale of the DL instability, $\tau \sim 1/kU_f$ is usually much longer than the characteristic time of the flame front inversion observed in experiments and simulations. At the same time, it should be stressed that already earlier simulations have shown the presence of a reverse flow in the burned gases behind the flame and a transverse velocity gradient along the flattened flame front near the sidewalls, as well as a slowdown of the central part of the flame surface. However, the resolution of simulations at that time was too low to distinguish causes from consequences, so the simulations do not confirm or exclude the participation of the DL instability in the inversion of the flame front. In addition, it should be noticed that the onset of flame front inversion cannot be explained simply by linear stability analysis of the DL instability. Dold and Joulin [27] hypothesized that the tulip flame has little or nothing to do with the deceleration of the gas velocity caused by a decrease in the surface area of the flame front. By means of numerical simulation of the modified Michelson–Sivashinsky equation they demonstrated that combined influence of the flame front curvature, geometric nonlinearity and the Darrieus–Landau instability is sufficient to produce the inversion of the flame front.

---

[1] Aspect ratio is often determined as D/L, but here we prefer to define aspect ratio as L/D.

An elegant theoretical explanation for the formation of a tulip flame was proposed by Matalon and Metzener [28, 29]. They derived a nonlinear evolution equation valid in the limit of weak thermal expansion, $q = QY_0 / c_p T_o << 1$, where $Q$ is the total heat of combustion, $Y_0$ is the initial mass fraction of the reactants. This equation describes the motion and the instantaneous shape of the two-dimensional flame front as a function of its mean position and contains a destabilizing term that results from the gas motion induced by thermal expansion and a memory term associated with vorticity generation. Numerical solutions to this equation demonstrated the development of convex and tulip flames. Since in the burned gas the flow is no longer potential because vortices can be produced by the curved flame front, the authors [29] emphasized that although the evolution equation contains a destabilizing term arising from the DL instability, it is the vortex motion that can lead to the formation of a tulip flame, depending on the initial conditions. The approach developed by Matalon and Metzener [28, 29] is the only analytical model developed in an attempt to explain the mechanism of tulip flame formation. However, the numerical simulations by Dunn-Rankin et al. [10] demonstrated the tulip flame formation in simulations with zero viscosity and without vortices. Since the formation of a tulip flame essentially depends on many parameters, such as aspect ratio, laminar flame velocity, the intensity of the ignition (the size of ignition spot), an analytical solution to the problem seems impossible.

It should be noted that the term "tulip flame" is often used for concave flames with a cusp pointing towards the burnt gas, when such shape of the flame develops as a result of different physical processes, e.g. the Darrieus- Landau instability [17, 18, 23, 27, 28, 29] or the flame-shock wave interaction [7]. To avoid confusion, we want to emphasize that the term "tulip flame" is used in this study to describe the purely gas-dynamic phenomenon of the flame front inversion from a convex to a concave shape, which occurs during an early stage of the flame propagation, when the flame was ignited near the closed end of the tube with non-slip walls and propagates to the opposite closed or open end, and when neither the DL instabilities nor the shock wave are involved.

Despite a large number of experimental, theoretical and numerical studies, there is still no unambiguous explanation of the physical mechanism responsible for the flame front inversion. Recent experimental studies by Ponizy, Claverie and Veyssière [1] were focused primarily on the flame front inversion and the mechanism of the tulip flame formation. The experiments were carried out with a stoichiometric propane-air mixture using cylindrical tubes closed at both ends with different aspect ratios, as well as for a tube closed at the ignition end with the opposite open end. A detailed study of gas dynamics in the upstream and downstream flows was carried out using a high-speed time resolved Particle Image Velocimetry (PIV) system based on a Nd:YLF two-cavity laser. The authors concluded that the inversion of the flame front is caused by pure gas-dynamic processes, which do not involve the intrinsic flame front instabilities (Darrieus–Landau, Rayleigh–Taylor, Richtmyer–Meshkov). They stressed that the initial "convex flame front has no influence on the phenomenon, but only the lateral parts of the flame front with radial expansion of burned gases are necessary". They found that the expansion of burned products deflects downstream and push the flame forward, and especially the gases burned near the wall are the main factors in the formation of a tulip flame. Inversion occurs by the action of two opposed flows: the reverse flow of the burned gases and the forward directed flow of unburned, previously accelerated gases.

The purpose of this paper is to study the dynamics and evolution of early stages of a laminar flame leading to the flame front inversion using high resolution 2D simulations of the fully compressible reactive Navier–Stokes equations. We show that, in agreement with experimental results [1], the physical mechanism of the tulip flame formation is indeed a purely gas-dynamic process. It is shown that rarefaction waves arising during the deceleration phase of the flame cause the reduction in the velocities of the unburned gas ahead of the flame, which (together with the reflected from the opposite end pressure waves), in turn, leads to the decrease in the flame front velocities uneven in the transverse (radial) direction, which explains the inversion of the flame front and the formation of a tulip flame in tubes with both closed ends and in half open tubes. Our study confirms the hypothesis of Guénoche [6] that rarefaction waves generated by the decelerating flame are a key factor explaining the formation of tulip flames.

## 3. Problem setup and basic equations

We consider a laminar stoichiometric hydrogen-air and methane/air flames ignited near the left closed end of the two-dimensional rectangular channel (tube) and propagating towards the opposite closed or open end. The computations solved the multidimensional, time-dependent, reactive compressible Navier-Stokes equations including molecular diffusion, thermal conduction, viscosity and chemical kinetics. The governing equations describing the flow are

$$\frac{\partial \rho}{\partial t}+\frac{\partial(\rho u)}{\partial x}+\frac{\partial(\rho v)}{\partial y}=0, \tag{1}$$

$$\frac{\partial(\rho u)}{\partial t}+\frac{\partial(P+\rho u^2)}{\partial x}+\frac{\partial(\rho u v)}{\partial y}=\frac{\partial \sigma_{xx}}{\partial x}+\frac{\partial \sigma_{xy}}{\partial y}, \tag{2}$$

$$\frac{\partial(\rho v)}{\partial t}+\frac{\partial(\rho u v)}{\partial x}+\frac{\partial(P+\rho v^2)}{\partial y}=\frac{\partial \sigma_{xy}}{\partial x}+\frac{\partial \sigma_{yy}}{\partial y}, \tag{3}$$

$$\frac{\partial(\rho E)}{\partial t}+\frac{\partial[(\rho E+P)u]}{\partial x}+\frac{\partial[(\rho E+P)v]}{\partial y}=$$
$$=\frac{\partial(\sigma_{xx}u+\sigma_{xy}v)}{\partial x}+\frac{\partial(\sigma_{yx}u+\sigma_{yy}v)}{\partial z}-\frac{\partial q_x}{\partial x}-\frac{\partial q_y}{\partial y} \tag{4}$$

$$\frac{\partial \rho Y_i}{\partial t}+\frac{\partial \rho u Y_i}{\partial x}+\frac{\partial \rho v Y_i}{\partial y}=\frac{\partial}{\partial x}\left(\rho D_i \frac{\partial Y_i}{\partial x}\right)+\frac{\partial}{\partial y}\left(\rho D_i \frac{\partial Y_i}{\partial y}\right)+\dot{\omega}, \tag{5}$$

$$P=\rho R_B T=\left(\sum_i \frac{R_B}{W_i}Y_i\right)\rho T=\rho T \sum_i R_i Y_i, \tag{6}$$

Here $\rho$, $u \equiv u_x$, $v \equiv u_y$, $T$, $P$, $E$, $Y$, $W_i$ are density, $x$, $y$ components of velocity, temperature, pressure, specific total energy, mass fraction and molar weight of species $i$. $R_B$ - is the universal gas constant. The viscosity stress and thermal energy flux are

$$\sigma_{xx}=2\mu\frac{\partial u}{\partial x}-\frac{2}{3}\mu\left(\frac{\partial u}{\partial x}+\frac{\partial v}{\partial y}\right), \tag{7}$$

$$\sigma_{yy} = 2\mu \frac{\partial v}{\partial y} - \frac{2}{3}\mu \left( \frac{\partial u}{\partial x} + \frac{\partial v}{\partial y} \right) \tag{8}$$

$$\sigma_{yx} = \sigma_{xy} = \mu \left( \frac{\partial v}{\partial x} + \frac{\partial u}{\partial y} \right). \tag{9}$$

$$q_x = -\kappa \frac{\partial T}{\partial x} - \sum \rho Y_i D_i h_i \frac{\partial Y_i}{\partial x} \tag{10}$$

$$q_y = -\kappa \frac{\partial T}{\partial y} - \sum \rho Y_i D_i h_i \frac{\partial Y_i}{\partial y} \tag{11}$$

The thermal properties for the fresh mixture and combustion products are taken with the temperature dependence of the specific heats, heat capacities and enthalpies of each species borrowed from the JANAF tables and interpolated by the fifth-order polynomials [30], though an ideal gas equation of state was found to be completely satisfactory. The transport coefficients were calculated from the first principles using the gas kinetic theory [31]. The viscosity coefficients for the gaseous mixture are

$$\mu = \frac{1}{2}\left[ \sum_i \alpha_i \mu_i + \left( \sum_i \frac{\alpha_i}{\mu_i} \right)^{-1} \right]$$

where $\alpha_i = \frac{n_i}{n}$ is the molar fraction, $\mu_i = \frac{5}{16} \frac{\sqrt{\pi \hat{m}_i kT}}{\pi \Sigma_i^2 \tilde{\Omega}_i^{(2,2)}}$ is the viscosity coefficient of $i$-species, $\tilde{\Omega}^{(2,2)}$ - is the collision integral which is calculated using the Lennard-Jones potential [31], $\hat{m}_i$ is the molecule mass of the i-th species of the mixture, $\Sigma_i$ is the effective molecule size. The thermal conductivity coefficient of the mixture is

$$\kappa = \frac{1}{2}\left[ \sum_i \alpha_i \kappa_i + \left( \sum_i \frac{\alpha_i}{\kappa_i} \right)^{-1} \right].$$

Coefficients of the heat conduction of i-th species $\kappa_i = \mu_i c_{pi} / \Pr$ can be expressed via the kinematic viscosity $\mu_i$ and the Prandtl number, which is taken $\Pr = 0.75$.

The binary coefficients of diffusion are

$$D_{ij} = \frac{3}{8} \frac{\sqrt{2\pi kT \hat{m}_i \hat{m}_j / (\hat{m}_i + \hat{m}_j)}}{\pi \cdot \rho \cdot \Sigma_{ij}^2 \tilde{\Omega}^{(1,1)}(T_{ij}^*)},$$

where $\Sigma_{ij} = 0,5(\Sigma_i + \Sigma_j)$, $T_{ij}^* = kT / \varepsilon_{ij}^*$, $\varepsilon_{ij}^* = \sqrt{\varepsilon_i^* \varepsilon_j^*}$; $\varepsilon^*$ are the constants in the expression of the Lennard-Jones potential, and $\tilde{\Omega}_{ij}^{(1,1)}$ is the collision integral similar to $\tilde{\Omega}^{(2,2)}$ [31].

The diffusion coefficient of i-th species is

$$D_i = (1 - Y_i) / \sum_{i \neq j} \alpha_i / D_{ij}.$$

A high-resolution simulation was used with minimum grid size, $dx \leq 20\mu m$, corresponding to 16 computational cells over the flame width. Since the tulip flame formation is expected to be a purely gas-dynamical process, a one-step Arrhenius chemical model was used in simulations, which results were compared with the results obtained in simulations with a detailed chemical model for hydrogen/air [32].

The 2-D direct numerical simulations were performed using the DNS solver, which used the fifth order weighted essentially non-oscillatory (WENO) finite difference schemes [33] to resolve the convection terms of the governing equations. The advantage of the WENO finite difference method is the capability to achieve arbitrarily high order accuracy in smooth regions while capturing sharp discontinuity. To ensure the conservation of the numerical solutions, the fourth order conservative central difference scheme is used to discretize the non-linear diffusion terms [34]. The time integration is third order strong stability preserving Runge–Kutta method [35].

## 4. Earlier stages of flame dynamics in a tube with no-slip walls

Before the development of a tulip flame shape, different stages of flame propagation can be distinguished. After the flame ignition at the closed end of the tube, the flame front quickly takes a nearly hemispherical shape. Clanet and Searby [15] have shown that the next stage is the formation of the finger shape flame, with the flame tip $X_{tip}$ accelerating as

$$X_{tip} \propto \exp(4\Theta U_f t / D), \tag{12}$$

where $D$ is the tube diameter (width in the 2D case), $\Theta = \rho_u / \rho_b$ is the ratio of the densities of unburned $\rho_u$ and burned, $\rho_b$ gases, and $U_f$ is the velocity of the laminar flame. This stage lasts a short time $t \approx D / 2\Theta U_f$ and it ends when the rear edge of the flame skirt touches the sidewalls. The expansion of the high temperature burned products between the flame and the closed end of the tube pushes the unburned gas towards the opposite end of the tube, thus creating an upstream flow ahead of the flame. A flame initiated near the closed end of the tube controls the upstream flow that forms ahead of the flame front, which leads to the flame acceleration compatible with the boundary conditions. From the conservation of the mass flux it follows that the velocity at the upstream flow is equal to $u_u = (\Theta - 1)U_f$, while the flame propagates with the velocity $U_{fL} = \Theta U_f$ relative to the tube wall and with the normal velocity of the laminar flame $U_f$ relative to the unburned gas flow. Because of the wall friction the velocity in the upstream flow is maximal at the tube axis and drops to zero at the tube walls. The inhomogeneous flow stretches the flame so that different parts of the flame front move at different speeds, and the flame front takes a shape similar to the velocity profile in the flow ahead of the flame. The flame sheet "repeats" to some extent the shape of the velocity profile in the upstream flow, remaining almost flat in the bulk (actually retaining the convex shape acquired earlier) with the trailing edges of the flame skirt extending backward in the boundary layer. Every point at the flame front moves relative to the unreacted mixture with velocity $U_f$ and simultaneously it is carried by the upstream flow with its local velocity $u_u^+(x,y)$ immediately ahead of this point. In the laboratory reference frame, the local velocity of the flame front at the point $(x,y)$ is[2]

$$U_{fl}(x,y) = U_f + u_u^+(x,y), \tag{13}$$

The velocity profile in the upstream flow ahead of the flame depends on the tube width. In a wide tube, the velocity is constant and parallel to the tube wall in the bulk and drops to zero within the boundary layer of thickness: $\delta_l \ll D$, while in a narrow tube the Poiseuille flow with a parabolic velocity profile develops in a short time $\sim D^2/100\nu$ [36, 37].

The stretched flame consumes fresh fuel over a larger surface area, which results in an increase in the rate of heat release per unit projected flame area. The increase in the rate of heat release due to stretching of the flame surface results in a higher volumetric burning rate, and a higher effective burning velocity based on the average heat release rate per frontal area of the flame sheet. A higher burning velocity results in an increase of the flow velocity ahead of the flame, which in turn leads to an increase in the flame front stretching and an increase in the burning speed. Thus, a positive feedback coupling is established between the upstream flow and the combustion wave velocity. Within the model of a thin flame front, the increase in the burning rate is proportional to the relative increase in the flame surface area. With accuracy $\delta_l/D \ll 1$ the flame surface grows linearly in time. Note, that in a relatively wide channel, the area within the boundary layer is the main contributor to the increase of the flame surface, which leads to an exponential increase of the flame velocity [36], similar to formula (12)

$$U_{fL} \propto \exp(\alpha \Theta U_f t / D), \tag{14}$$

where $\alpha$ is a dimensionless coefficient of the order of unity. Note that, when a flame is ignited (e.g. by an electrical spark), the flame front is not flat, and during, the first stage it has a finger-like shape and then a convex shape. However, the velocity profile in the upstream flow is flat in the bulk already in a short distance ahead of the flame tip.

---

[2] It should be noted that, strictly speaking, Eq. (13) is valid for a steady flow. In the case of unsteady flow this equation should be considered as the tendency of the dependence of the local velocity of the flame front on the velocity of the unburned gas.

After the rear edges of the flame skirt touched the sidewalls, the flame skirt begins to stretch along the side walls, the angle between the flame skirt and the side wall decreases, and the rear part of the flame skirt becomes almost parallel to the sidewalls. Shortly thereafter, this part of the flame skirt touches the side wall and quenches, which leads to a sharp decrease in the flame surface area and as a consequence, to a sharp decrease in the average speed of the combustion wave[3]. The decrease of the flame surface area leads to the decrease in the average speed of combustion wave and to the decrease in the velocity of the burned gas expansion, which can be roughly estimated using a one-dimensional model [15] as, $\langle u_b \rangle = \dfrac{2L}{D} \Theta U_f$, where $L$ is the distance from the flame tip to the rear end of the tube. The most important effect of the flame deceleration is that the decelerating flame begins to generate rarefaction waves, which reduce the flow velocity ahead of the flame. It should be noticed that, unlike a stationary flame, the flow with the accelerating flame is not isobaric. During the accelerating stage, the pressure rises at about the same rate as the flame velocity, producing pressure waves and pressure gradients that support the acceleration of the flow ahead of the flame. The effect of the rarefaction waves leads to a decrease in the flow velocity of the unburned gas and, apparently, in the establishing of an inverse pressure gradient. It should be noted that in the case of a relatively short tube with both ends closed, and a fast flame, the compression waves reflected from the end of the tube opposite to the ignition end, also contribute to the decrease in the flow velocity ahead of the flame, enhancing the effect of rarefaction waves. For a fast flame propagating in a tube with both ends closed, this can lead to the formation of an inversed flow of the unburned gas directed towards the flame.

---

[3] Depending of the size of the ignition kernel and the tube width, the flame skirt may contact the sidewalls during "finger shape" phase.

The reduced velocity of the flow ahead of the flame leads to an increase in the width of the boundary layer and a corresponding change in the velocity profile ahead of the flame, which determines the specific shape of the tulip petals. A few comments should be made about a tulip flame formation depending on a channel width. The thickness of a laminar boundary layer in the upstream flow is $\delta_l \approx 5X/\sqrt{\mathrm{Re}}$, where $\mathrm{Re} \approx u_u X/\nu \approx X(\Theta-1)U_f/\nu$ is the Reynolds number, $X$ is the coordinate along the tube. The expression $\delta_l \approx 5X/\sqrt{\mathrm{Re}}$ was obtained for the case of a steady flow. For unsteady flow, $\delta_l \approx X/\sqrt{\mathrm{Re}}$ is in better agreement with the observed thickness of the boundary layer. The mass of the unburned mixture passing through the tube cross-section is the same, but the thickness of the boundary layer increases with $X$. Therefore, the inner part of the flow, where the velocity is uniform, accelerates with the increase of $\delta_l$. In general, this leads to the development of a Poiseuille flow with a parabolic velocity profile. The Poiseuille flow is formed when the boundary layer thickness becomes equal to half of the tube width. An estimate of the distance that the flame travels and the corresponding time of establishing a parabolic velocity profile is: $X_P \sim \Theta U_f D^2/100\nu$ and $t_P \sim D^2/100\nu$ [36]. A more detailed analysis [38, 39] leads to a slightly different numerical coefficient: $X_P \approx \Theta U_f D^2/104\nu$, and $t_P \approx D^2/104\nu$. Using the relation $U_f L_f \sim \nu$, we can write these formulas in the form: $X_P \approx \Theta D^2 / 100 L_f$ and $t_P \approx D^2 / 100\nu$. For example, in a hydrogen/air mixture ($L_f = 0.035\,\text{cm}$, $\Theta = 7.8$) the Poiseuille flow in the tube of width $D = 1\,cm$ is established after a rather long time $t_P \approx 50\,\text{ms}$ compared to the characteristic times of the flame front inversion $t_{tulip} \approx 1.0\,ms$ or the transition to detonation $t_{DDT} \approx 2.0\,\text{ms}$. On the contrary, in narrow channels, $D \leq 1\,\text{mm}$, the Poiseuille flow is established in a short time $t_P \sim 0.1\,\text{ms}$. Note that in narrow tubes, $D \leq 1\,\text{mm}$, where the Poiseuille flow is established at the earliest stage of flame propagation, the flame accelerates without a deceleration stage, and this is the reason why the formation of tulip flame in narrow tubes has never been observed experimentally. This confirms the assumption that the boundary layer plays a key role in the tulip flame formation.

## 5. Results of numerical simulations

### 5.1. Hydrogen/air flame in a tube with both closed ends; a one-step Arrhenius model

The numerical simulations of the stoichiometric premixed hydrogen/air flame presented in this section were performed for a tube with both closed ends, with a tube width $D = 0.6\,\text{cm}$, and a tube length $L = 8.4\,\text{cm}$; the aspect ratio $L/D = 14$. The flame was ignited near the left closed end and propagates to the opposite closed end. It is known that a one-step Arrhenius chemical model is computationally much easier for multidimensional simulations if it can satisfactory reproduce flame dynamics. Since, according to experiments [1], the formation of a tulip flame is expected to be purely gas-dynamical process, a one-step Arrhenius chemical model was used in simulations. The parameters of a one-step chemical model for a premixed stoichiometric hydrogen/air flame with the Lewis number equal to unity were taken the same as in [40]. It should be noted that the laminar hydrogen/air flame velocity obtained for a one-step model [4] is $U_f = 3.0$ m/s, while the detailed chemical model and experiments give the laminar flame velocity $U_f = 2.36$ cm/s. Comparison of simulations with one-step and detailed chemical models will be discussed in Sec. 5.4.

Figure 1 shows the calculated time evolution of the local speed of different points at the flame front during the tulip shape formation: at the channel axis ($y = 0$), $U_{fL}(0)$, near the sidewall at $y = 0.16$ cm, $U_{fL}(0.16\text{cm})$, the average combustion wave speed $S_f$, and the flame surface area (length) $F_f$. The combustion wave velocity (the effective burning velocity) is defined as

$$S_f = \frac{1}{D}\int_0^D U_{fL} \cdot \sqrt{1+(\partial x/\partial y)^2}\, dy,$$

where $U_{fL}(x, y)$ is the $x$ component of the local speed of the flame front at the point (x, y) on the flame surface in the laboratory reference frame.

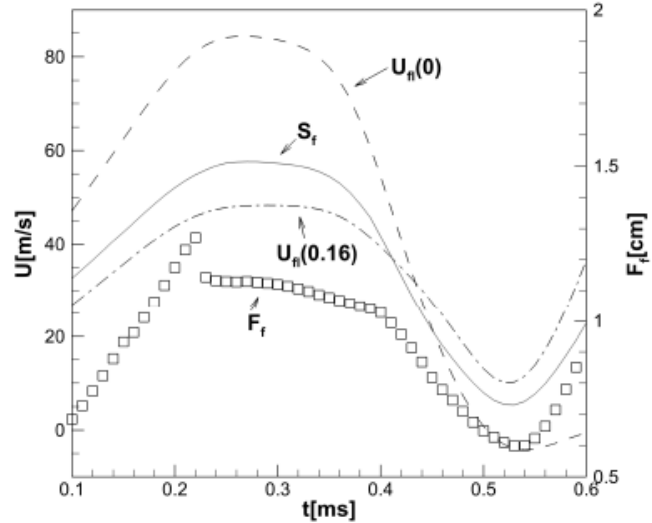

Figure 1. The time evolution of combustion wave velocity $S_f$ (solid line), local x-component of the flame front velocities $U_{fL}(x,0)$ at y = 0 and at y = 0.16 cm, and the flame surface area (length) $F_f$ shown by squares. The tube width $D = 0.6$ cm, $L/D = 14$, both ends closed.

In Fig.1 can be seen the initial exponential increase in the combustion wave velocity and the velocity of the flame tip ($U_{fl}(y=0)$), followed by a subsequent sudden sharp decrease of the flame surface area (length) at $t \approx 0.22ms$, and the transition to the deceleration stage, during which the flame surface area, the average flame speed $S_f$ and local velocities of different parts of the flame front continue to decrease. The growth of the flame speed during the acceleration stage looks as linear in Fig. 1 because of the short time of the acceleration stage. After $t \approx 0.22ms$ the rarefaction wave created by decelerating flame begin to decrease the flow velocity ahead of the flame. After $t \approx 0.42ms$, the compression wave created by the flame during the acceleration phase and reflected form the opposite end of the tube returns to the flame front and additionally enhances the decrease in the flow velocity ahead of the flame. As a result, the speed of the flame front along the line $y = 0.16\,cm$ closer to the side wall becomes equal and then exceeds the average flame speed, $U_{fl}(x,0.16) \geq S_f$. This indicates the beginning of the flame front flattering, which continues after 0.46ms, when $U_{fl}(x,0.16)$ exceeds the average flame velocity and the flame front velocity at the centerline $U_{fl}(y=0)$. After 0.46ms, the flame front velocity along centerline becomes less than the average velocity of the flame and the time interval $0.42ms < t < 0.47ms$ corresponds to the maximum deceleration of the flame. After $t \approx 0.47$ms the inversion of the flame front and the development of the tulip shape of the flame begin to be more and more noticeable.

Figure 2 shows the velocity profiles in the unburned gas, $U_+ = u_u(x_{f+}, y_f)$, immediately ahead of the flame at several selected time instants. It is seen that the velocity ahead of the flame front is largely reduced by the rarefaction waves at and near the tube axis, $y = 0$. Accordingly, the local velocity of various points on the flame front $U_{fl}(x,y) = U_f + U_+(x_{f+}, y_f)$ decreases more at and near the central line and less near the sidewall. After 0.55ms the velocity of the unburned gas is maximum at the line $y = 0.16 cm$, where the tulip petal is formed. It should be emphasized that due to the decrease in the velocity of the unburned gas, the thickness of the boundary layer in the flow ahead of the flame increases. In particular, for the case considered in this section, the reflected compression waves generated during the acceleration phase, together with the rarefaction waves generated by the flame during the deceleration phase, create a reversed flow ahead of the flame. It is seen in Fig. 2 that the velocities in the flow ahead of the flame are negative and directed towards the flame for $0 < y < 0.1 cm$ and for $0.22 cm < y < 0.3 cm$. During the inversion of the flame front, which begins at $t \approx 0.49$ ms, the thickness of the boundary layer becomes $\delta_1 \approx 0.05$ cm, which agrees with Fig. 2, and is about of the thickness of the tulip petal from the side closer to the wall.

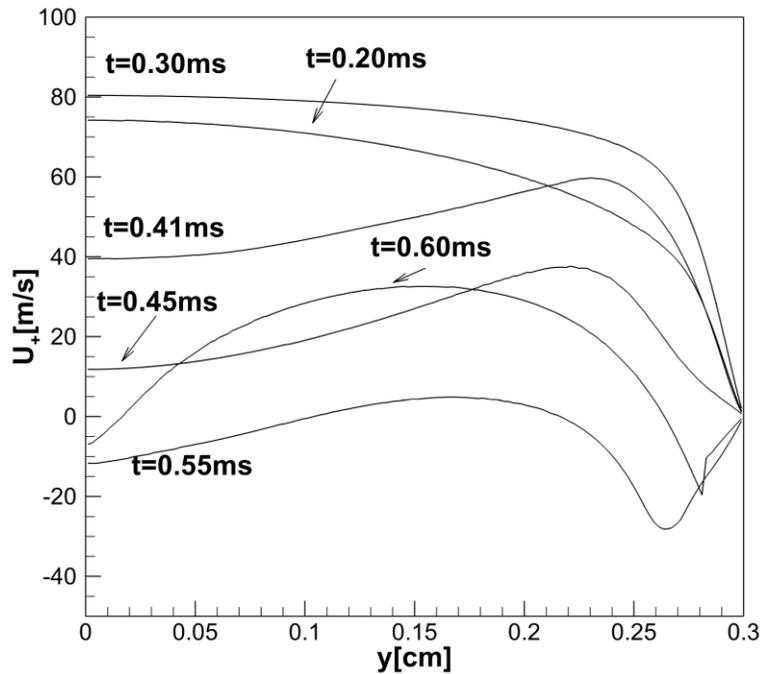

Figure 2. Velocity profiles of the unburned gas immediately ahead of the flame at selected times.

Figure 3 shows the velocity profiles of different parts of the flame front after the beginning of the deceleration phase and until the inversion of the flame front occurs. It is seen that the velocities of the central part of the flame front near the axis ($0 \leq y \leq 0.1 cm$) are exposed to the most intense reduction, while the decrease in the velocities of the flame front segment $0.1 cm < y < 0.2 cm$ is essentially less with the local velocity of the flame front at $y = 0.16 cm$ undergoes a minimal reduction. This is the location where the tulip petal will be finally formed.

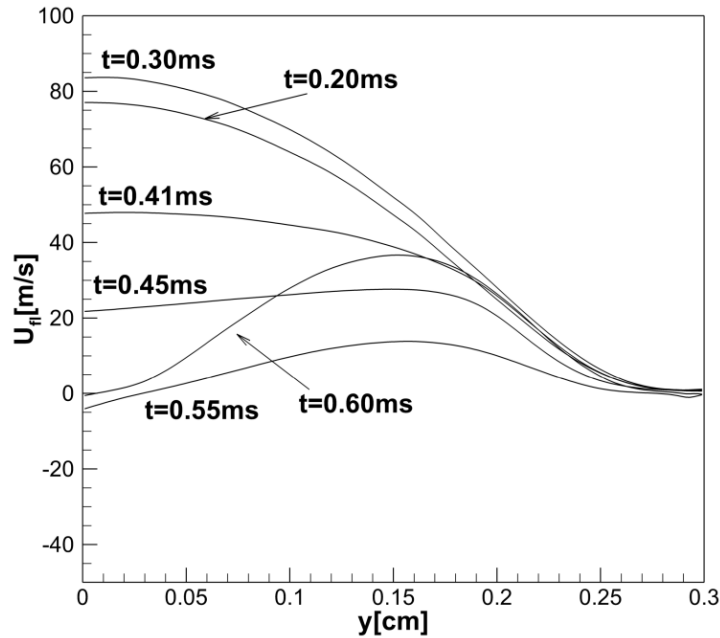

Figure 3. Profiles of the x-component velocities of the flame front at selected times.

The change in the local velocities of the flame front, leading to the formation of a tulip- flame, is associated with rarefaction waves created in the unburned mixture by the decelerating flame, which can be enhanced by the pressure waves reflected from the opposite end if the tube is not too long. The rarefaction waves travel to the right and, after reflection from the closed end of the tube, return to the flame front, partly weakening the effect of the original rarefaction wave. However, this is compensated by the reflected pressure wave. The flow velocity of the unburned gas ahead of the flame is stronger reduced near the tube axis, at $y = 0$. In this particular case of a relatively fast flame and a relatively short tube, a reverse flow of unburned gas towards the flame was established.

Figure 4 shows the computed schlieren images (density gradient) and streamlines at selected time instants during the tulip flame formation. At the first images one can see the sequence of compression waves generated by the flame during ignition and the accelerating stage.

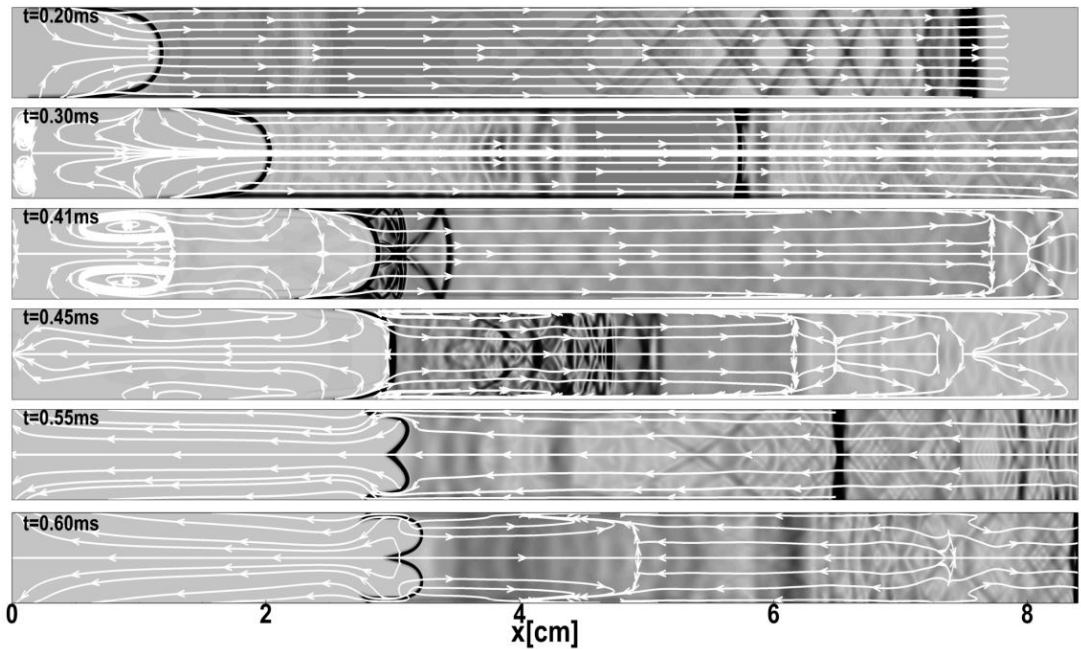

Figure 4. Sequences of calculated "schlieren images" and streamlines during the tulip flame formation for the condition of Figs. 1-3.

The pressure waves propagate towards the opposite closed end of the tube and after reflection from the opposite closed end of the tube and from the side walls they return to the flame almost at the same time with the rarefaction wave reflected from the ignition end. Before 0.4ms, the shape of the flame is still convex with the flame cusp directed towards the unburned gas. The flame front begins to flatten after about 0.42ms, when the velocities in the flow ahead of the flame have already been significantly reduced by the rarefaction waves together with reflected pressure waves. At the later time, at around 0.50ms the flame front begins to invert and after 0.55ms the tulip shape is already formed and well seen at the last schlieren image at 0.6ms. In this particular case of the "fast" flame and aspect ratio $L/D=14$ the rarefaction waves and reflected pressure waves are strong enough to invert the flow velocities ahead of the flame towards the flame. The schlieren image shows a vortex in the burnt gas at the time 0.41ms. A closer look shows that the vortices formed in the burnt gas subsequently disappear and do not affect the formation of the tulip shape.

Apparently, the vortices are formed as a result of the interaction of the gas flow reflected from the side wall to the tube axis and the flow along the tube the centerline of behind the flame front, which are adjusted in such a way as to satisfy the boundary conditions at the tube walls. In this case the vortices are not related to the intrinsic flame front instabilities as this was suggested in the theoretical model by Matalon and Metzener [28, 29], which explains the tulip flame formation by the vortices behind the flame, which drive and invert the flame front into a tulip shape. It does not appear that vortices play an important role, if any, in the flame front inversion. Note, that while the velocity of the burned gas was positive during the phase of the flame acceleration, it begins to change sign during the deceleration phase, and after 0.45ms the velocity in the flow behind the flame become negative, directed towards the ignition end.

Figure 5 shows the pressure and rarefaction waves during the acceleration (0.2ms) and deceleration phase, when the tulip flame is formed, as well as the change in the flow velocity along the centerline of the tube in the unburned gas ahead of the flame and in the burned gas behind the flame. The compression wave reflected from the right end of tube approaches the flame at 0.41ms and reduce the velocity in the unburned gas immediately ahead of the flame. In this case, the direction of the flow velocity ahead of the flame and behind the flame is inverted after the formation of the tulip flame and velocity becomes everywhere negative. After 0.6ms, when the tulip flame has formed and the tulip petals begin to stretch along the wall, increasing the flame surface area, the flame is accelerated and the flow immediately ahead of the flame will change direction again to positive.

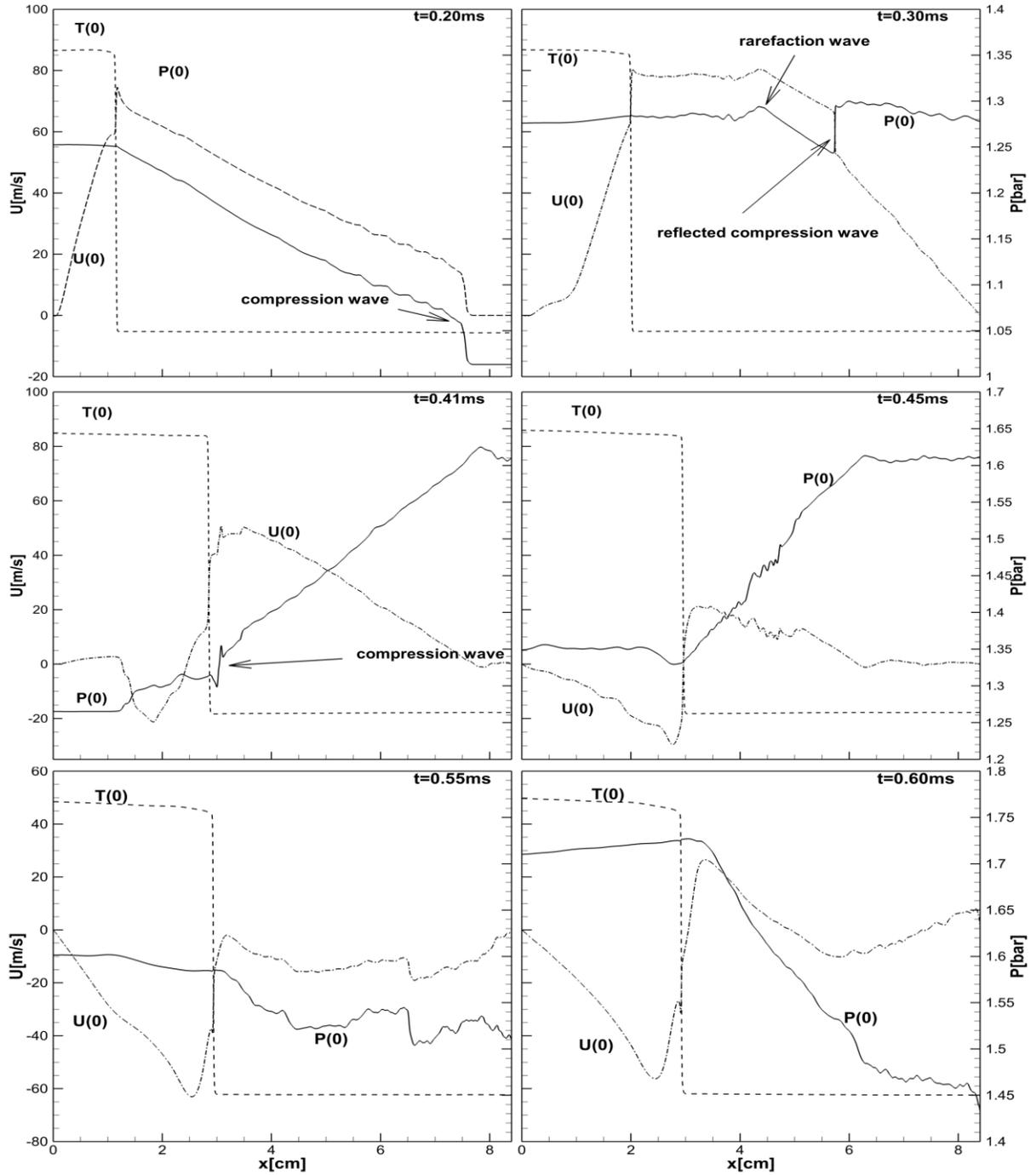

Figure 5. Pressure (solid line), temperature (dashed line) and flow velocity (dashed dotted lines) at the tube axis $y = 0$ during the tulip flame formation in hydrogen/air mixture.

## 5.2. Methane/air flame in a tube with both ends closed; a one-step Arrhenius model

Scenario of the flame front inversion in the case of a low-velocity flame is, in general, analogous to the scenario of the tulip flame formation for a fast flame in a tube with both closed ends. As an example, the formation of a tulip flame was modeled for the slowly reactive methane-air mixture with the flame propagating in a tube with both closed ends and the same width and length as in Sec. 5.1. The parameters of the one-step Arrhenius model for a methane/air flame were taken the same as in [41] with the laminar methane/air flame velocity $U_f = 0.38\,m/s$.

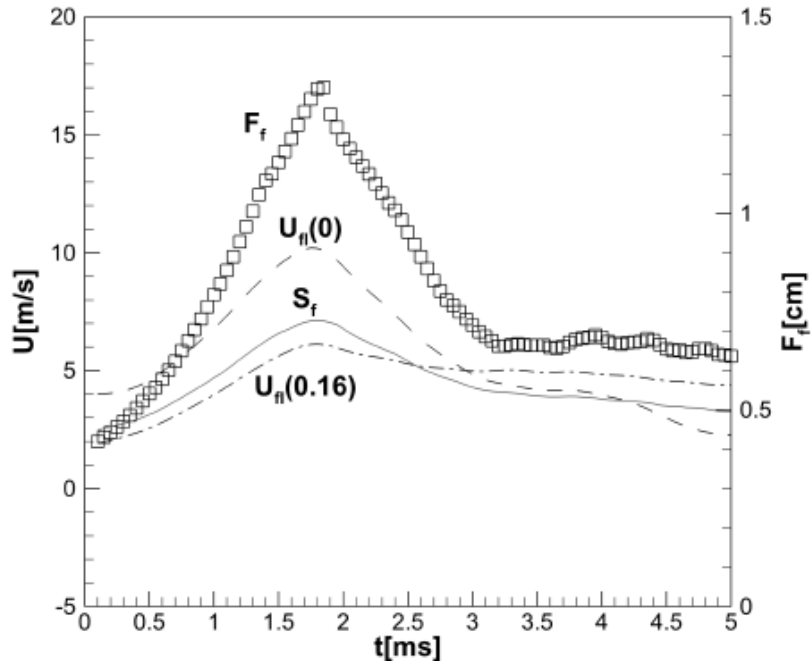

Figure 6. The time evolution of the methane/air flame surface area $F_f$, the average flame speed $S_f$ (solid line), the local x-component of the flame front velocities $U_{fL}(0)$ at $y=0$ and at $y=0.16\,cm$. The tube with both ends closed, $D=0.6\,cm$, $L/D=14$.

Figure 6 shows the time evolution of the flame surface area $F_f$ before and after extinguishing the rear part of the flame surface on the sidewalls, as well as the corresponding time evolution of the local velocities of the flame front at the channel axis, $U_{fL}(y=0)$ and along the sidewall at $U_{fL}(x, y=0.16\,cm) \equiv U_{fL}(0.16)$, and the average combustion wave speed $S_f$. It is seen that deceleration stage begins after 1.8ms, when extinguishing the rear parts of the flame skirt at the side walls, the flame surface area, the average flame velocity and the local velocities of the flame front $U_{fL}(0)$ and $U_{fL}(0.16)$ decrease. After 2.5ms the velocity of the flame front along the line $y = 0.16\,cm$ exceeds the average flame speed, which means that the flame front becomes almost flat. After 3ms the local velocity of the flame front $U_{fL}(0.16)$ exceeds the flame front velocity at the central line, $y = 0$, and this is the beginning of the flame front inversion. Finally, after 4.2ms, the tulip shape of the flame is formed: the velocity of the flame velocity at the central line becomes the lowest velocity, the side parts of the flame run forward with the central part trailing behind. The main differences are in the larger characteristic time scales due to the fact that the speed of the methane/air flame is much lower than the speed of the hydrogen/air flame, and a noticeably smaller the rate of the decrease of the flame the surface area and the velocities of the flame front.

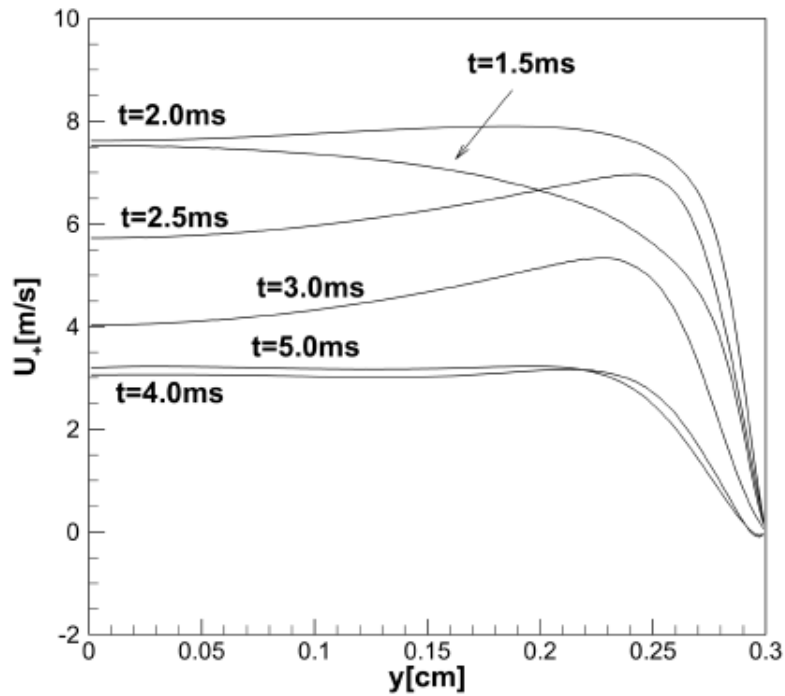

Figure 7. The velocity profiles in the unburned gas immediately ahead of the methane/air flame at the selected times.

Figure 7 shows the velocity profiles in the unburned gas immediately ahead of the flame front at the selected times during the tulip flame formation. It is seen that the decrease in the velocity of the unburned gas is not such dramatic as in the case of a hydrogen/air flame. It is stronger at and near the axis of the tube, but the velocity at the axis remains positive and does not inverted as in the case of a hydrogen/air flame. In this case the rarefaction and pressure waves are weaker than in the case of the fast hydrogen/air flame.

Figure 8 shows the x-components of the velocity of different parts of the methane/air flame front at selected times. Since the decrease in the speed of the unburned gas immediately ahead of the flame is maximal on the tube axis, the speed of the central part of the flame front also decreases more than the speed of its parts located closer to the wall. The evolution of the local velocities of the flame front leads to the maximum local velocity of the along the line $y = 0.16\,\text{cm}$, where tulip petals are formed.

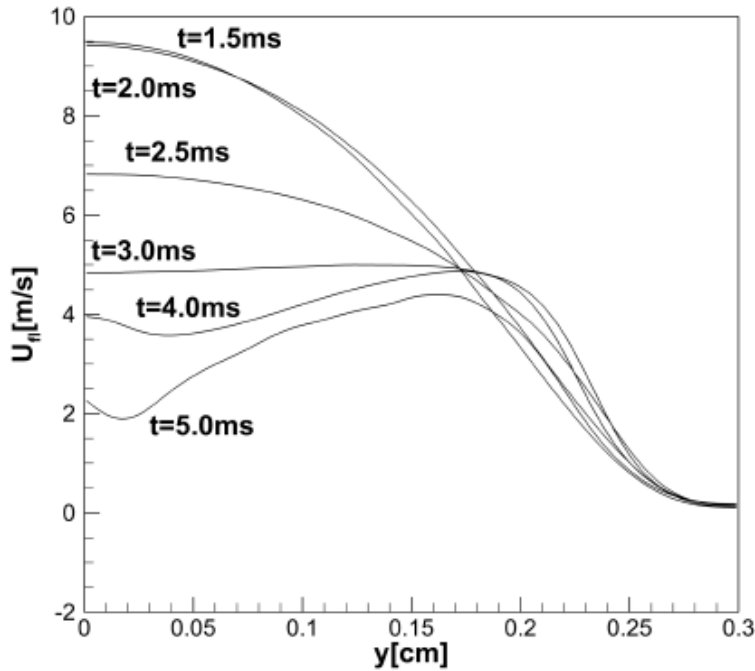

Figure 8. Local velocities of the methane/air flame front for the conditions in Fig. 7.

Figure 9 shows the calculated schlieren images and streamlines at the same selected times, as in Figs. 7, 8. It is seen that the flame front begins to flatten after 3ms, and it begins to invert after 3.5ms, which are close to the time when the weak pressure wave reflected from the right end approaches the flame and helps to reduce the velocity of the unburnt gas ahead of the flame.

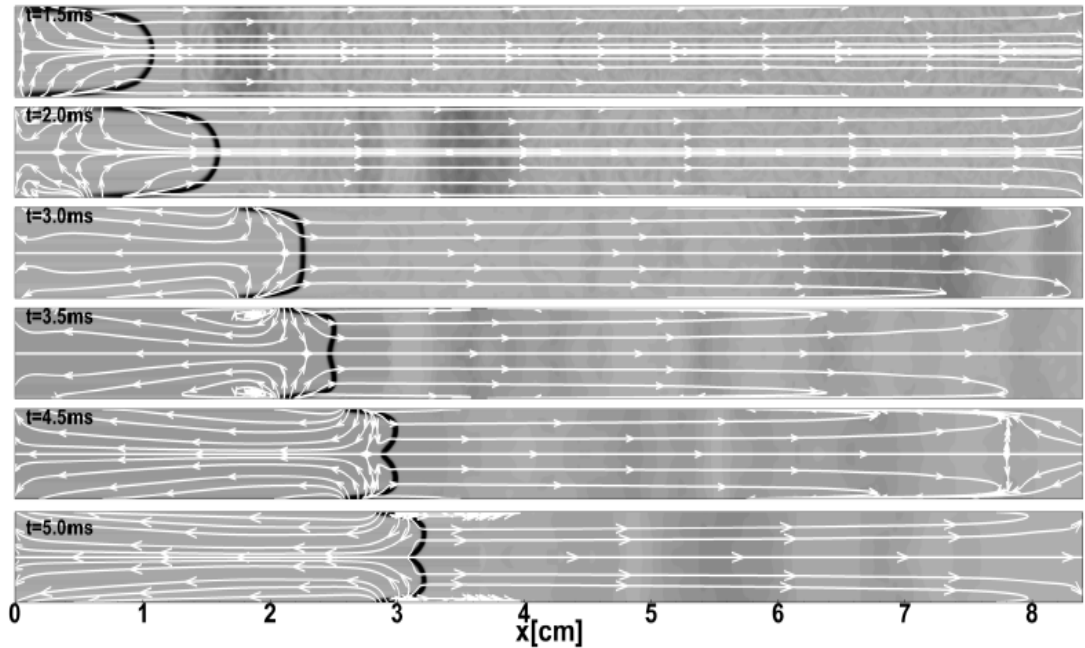

Figure 9. Sequences of schlieren images and streamlines during the tulip flame formation in a methane/air flame.

The schlieren images also show the appearance of vortices at 3.5ms, during the beginning of the flame front inversion. However, vortical structures do not develop and these vortices disappear and are not seen in schlieren images at later times. The vortices that arise in the burnt gas are apparently not directly related to the mechanism of the tulip flame formation. In the case of a hydrogen/air flame, the vortices in Fig. 4 appear in the schlieren image at 0.41ms, just before the flame front begins to flatten, and they disappear in subsequent schlieren images. In the case of a methane/air flame the vortices in Fig. 9 appear in the schlieren image at 0.35ms, when the flame front already became planar and, in fact, a tulip shape is developing. As in the case of the tulip shape formation in a hydrogen/air flame, the thickness of the boundary layer in the flow ahead of the flame increases due to the decrease in the velocity of the unburned gas mixture, which is rather low, $U_+ \approx 3 m/s$, at the time of the flame front inversion. At this time the thickness of the boundary layer is $\delta_l \approx 0.055$ cm which is close to the thickness of the tulip petal from the side closer to the wall.

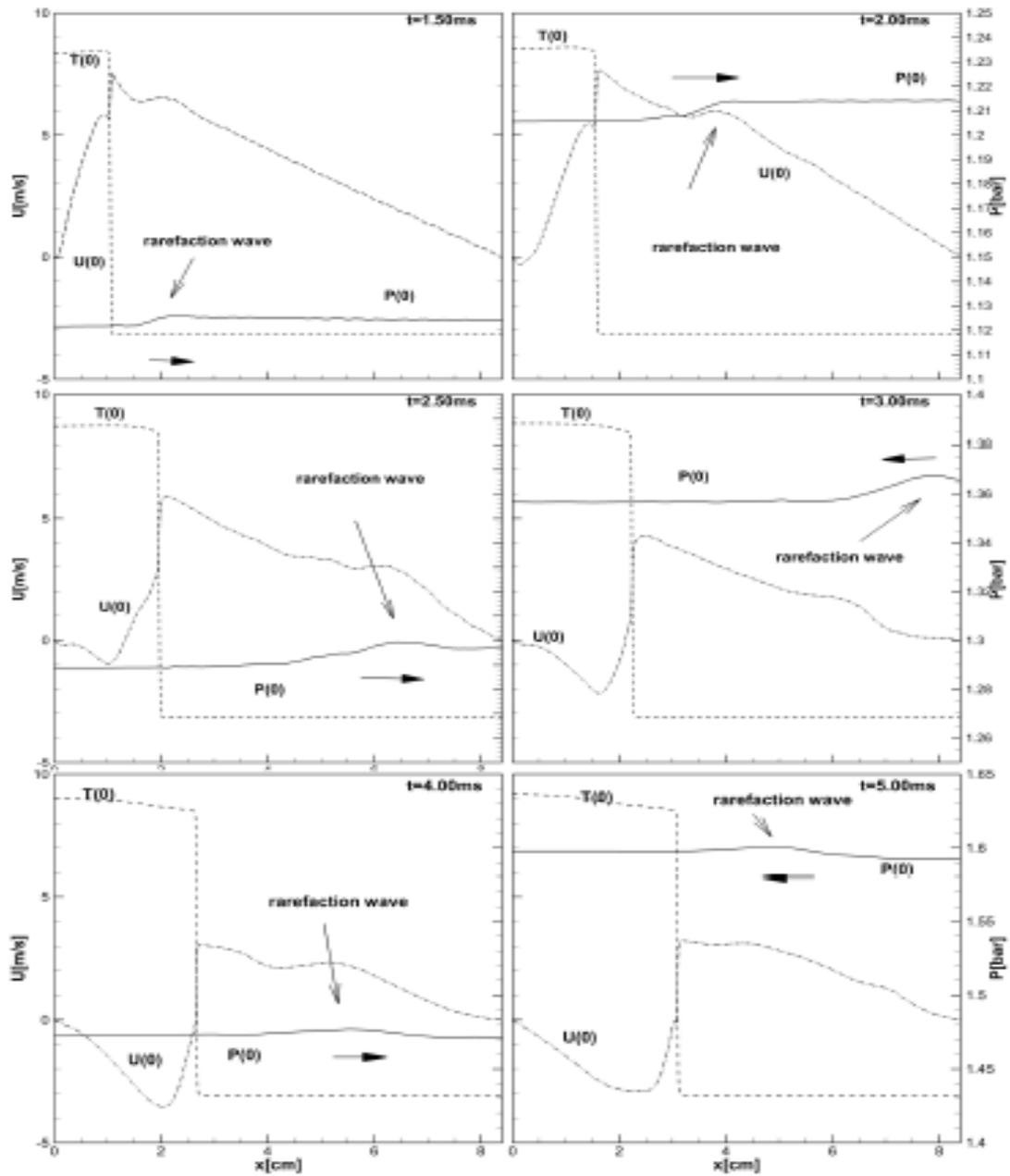

Figure 10. Pressure waves (solid line), temperature (dashed line) and flow velocity (dashed dotted lines) at the tube axis during the tulip flame formation in methane/air.

It should be noted that during the acceleration stage, the flow velocity ahead and behind the flame are positive and directed in the direction of the flame propagation. After the flame has acquired a tulip shape, the velocity in the flow ahead of the flame can be positive as in the case of a slow flame (Fig. 9), or can change direction to the negative as in the case of a hydrogen/air flame (Fig. 4). However, when the flame has developed a tulip shape, the flow velocity behind the flame always changes towards the ignition end of the tube for both cases of slow and fast flames.

Figure 10 shows rarefaction waves, pressure waves and the flow velocity along the centerline of the tube in the unburned gas ahead of the flame and in the burned gas behind the flame during the tulip flame formation. Compression waves reflected from the right end of the tube in the case of a slow methane/air flame can reach the flame, but they are weak and do not affect the formation of the tulip flame.

**5.3. Hydrogen/air flame in a half open tube; a one-step Arrhenius model**

The scenario of the formation of a tulip flame in the case of a hydrogen/air flame ignited at the closed end of the tube and propagating to the opposite open end is similar to the case of a slow or a fast flame in such long tubes that the time required for pressure waves reflected from the opposite end of the tube is too long to return and enhance the rarefaction wave effect. Simulations were performed for a stoichiometric hydrogen/air flame propagating from the closed end of the tube of width $D = 0.8\,\text{cm}$ and length $L = 13.6\,\text{cm}$ to the opposite open end.

Figure 11 shows the evolution of the flame surface area $F_f$, the average speed of the combustion wave $S_f$, and the local velocities of the flame front along the tube axis $y = 0$ and close to the side walls, at $y = 0.2\,\text{cm}$. The deceleration stage begins at 0.4ms when the flame surface area decreases due to the extinction of the rear part of the flame skirt. The velocity of the flame front near the side wall, at $y = 0.2\,\text{cm}$, exceeds the average velocity of the flame after 0.65ms.

After 1.33ms the velocity near the side wall $U_{fl}(0.2)$ exceeds the velocity of the flame front $U_{fl}(0)$ at the center line. This is the beginning of the inversion of the flame front. After 1.6ms the velocity of the flame front $U_{fl}(0)$ at the center line becomes less than the average speed of the combustion wave, which means that the flame front has acquired a tulip shape.

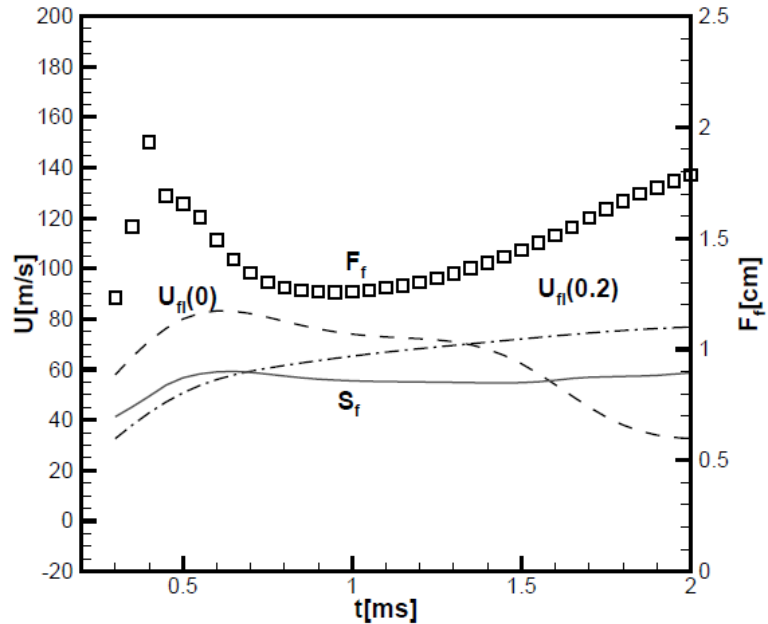

Figure 11. The evolution of the flame surface area $F_f$, the combustion wave velocity $S_f$ (solid line), and the local velocities of the flame front at $y = 0$ and $y = 0.2\,\text{cm}$.

The evolution of x-component of the flow velocity profiles immediately ahead of the flame at selected time instants is shown in Figure 12. As in the case of a tube with both closed ends, the rarefaction wave more strongly reduces the flow velocity ahead of the flame near the tube axis, thereby creating conditions for the inversion of the flame front. However, since there are no pressure waves reflected from the opposite end of the tube, the effect is not such strong as in the case of a tube with both closed ends.

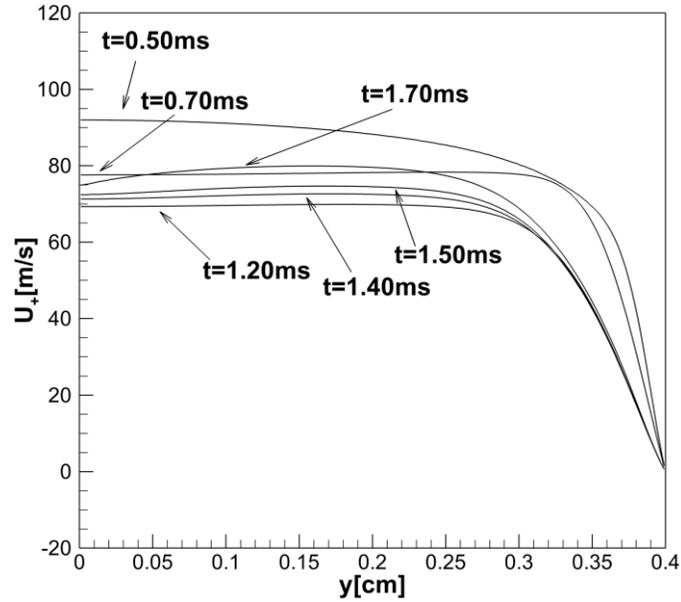

Figure 12. x-component of the flow velocity profiles ahead of the hydrogen/air flame at selected times.

Figure 13 shows the change in the velocity profiles of the flame front at selected times corresponding to the change in the flow velocities immediately ahead of the flame front. It is seen that the flame velocity at the centerline is exposed to a maximum decrease, while the minimum decrease in the flame front velocity occurs further from the center line at $0.1 cm < y < 0.2 cm$.

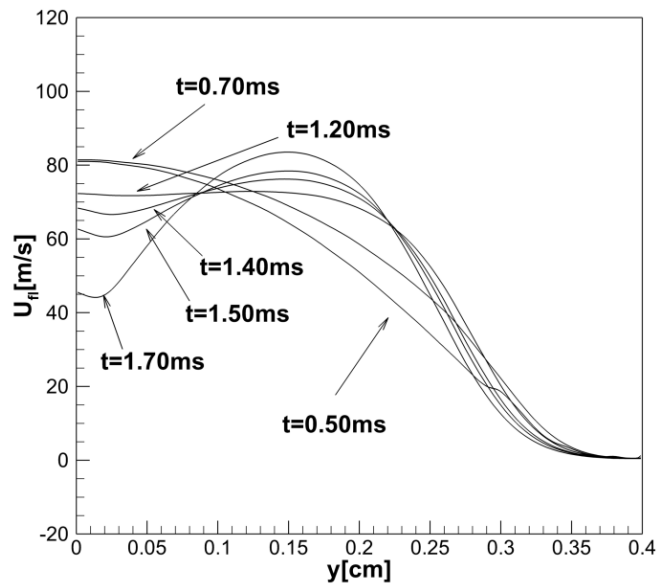

Figure 13. The evolution of x-component of the flame front velocity profiles at selected times.

Figure 14 shows a sequence of schlieren images during the formation of a tulip flame. The streamlines shown in Fig. 14 demonstrate the change in the direction of the flow velocity behind the flame after the beginning of flame front inversion, which is typical for all cases of the tulip flame formation. In the case where the influence of rarefaction waves is amplified by the reflected compression wave, which occurs in the case of a fast flame in a tube with both ends closed, the speed of the burnt gas changes direction towards the igniting end of the tube everywhere behind the flame. In the case of a half open tube, the velocity immediately behind the flame remains positive but it changes direction further behind the flame.

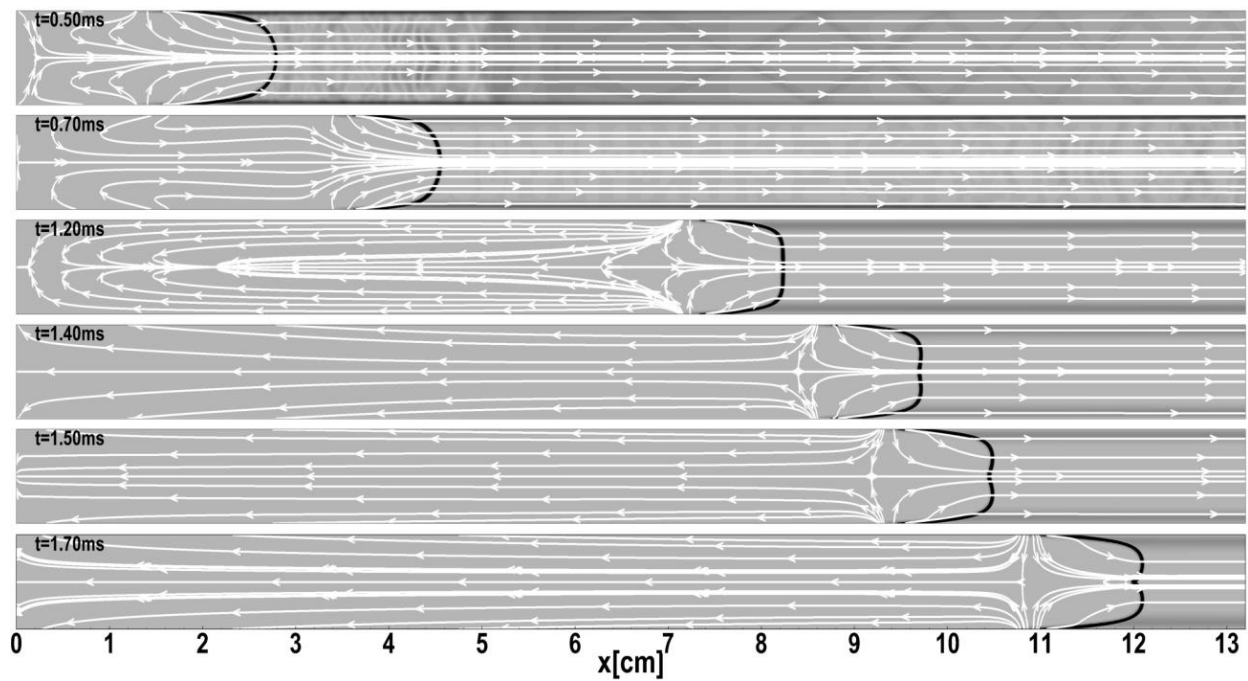

Figure 14. Sequences of schlieren images and streamlines during the tulip flame formation in a hydrogen/air flame.

The pressure profiles and rarefaction waves created by the flame during the deceleration stage, as well as the flow velocity along the centerline of the tube in the unburned gas ahead of the flame and in the burned gas behind the flame are shown in Figure 15.

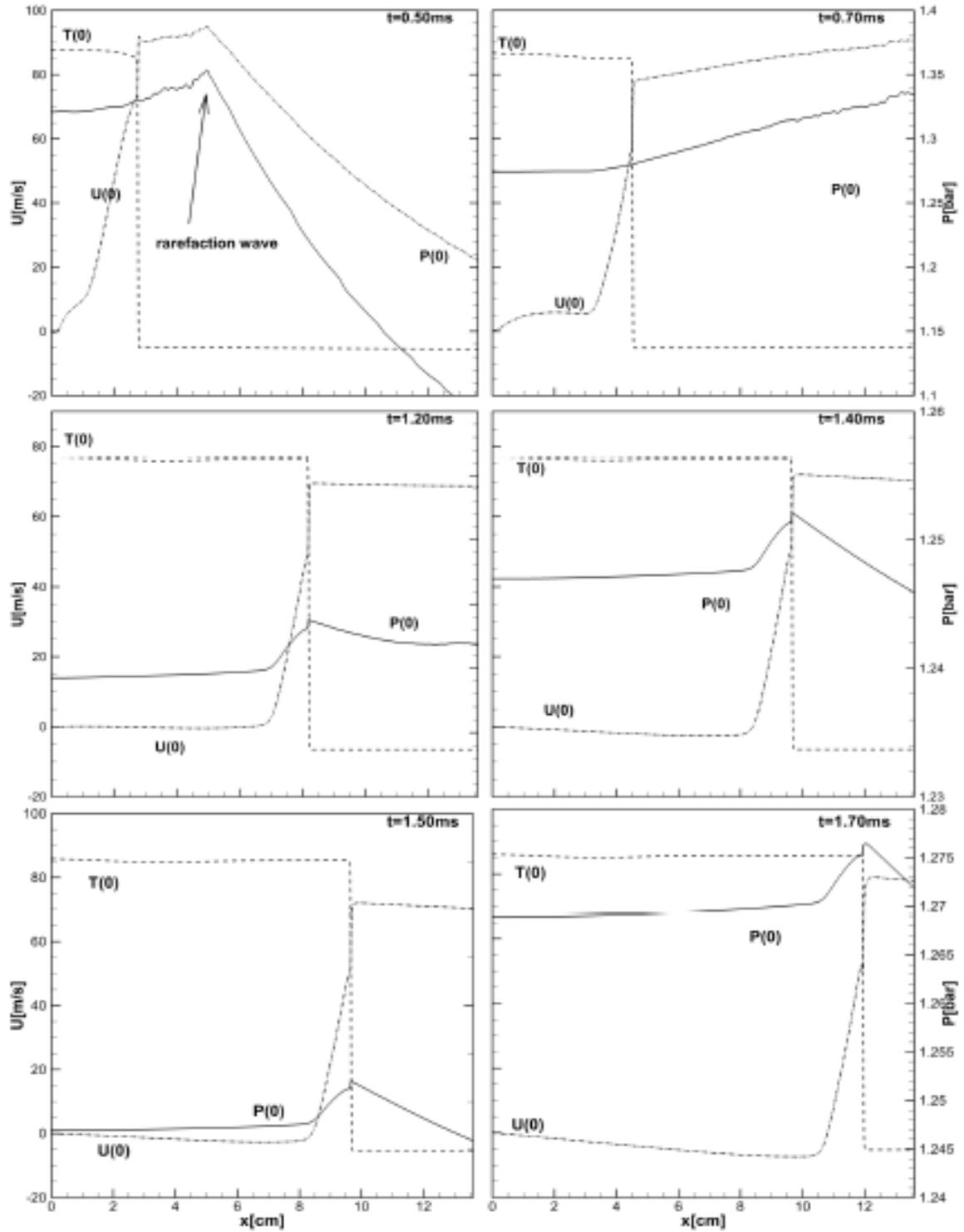

Figure 15. Profiles of pressure (solid line), temperature (dashed line) and flow velocity (dashed-dotted line) at the centerline of the half-open tube during the tulip shape formation of the hydrogen/air flame.

### 5.4. Whether a chemical model affects the formation of a tulip flame

To verify whether the choice of a chemical model influences the mechanism of a tulip flame formation, modeling of a tulip flame using the Arrhenius one-step chemical model was compared with modeling of a tulip flame using a detailed chemical model [32]. The laminar velocity of the hydrogen/air flame for the parameters of the one-step model [40] is $U_f = 3.0\,\text{m/s}$, which is higher than the flame velocity $U_f = 2.36\,\text{m/s}$ obtained with the detailed model [32] and in experimental studies. To eliminate the effect of the difference in velocities, the pre-exponential factor $A$ in the one-step model $\dot{\omega} = A\rho Y \exp(-E_a/R_B T)$ was slightly modified to make the laminar flame speed equal to $U_f = 2.3\,\text{m/s}$. The simulations were performed using a detailed chemical model [32] for the tube with both ends with the same sizes as in Section 5.1 and compared with simulations with a one-step chemical model in Sec. 5.1.

Figures 16 shows the change in the flame surface area and the average flame velocity obtained in simulations using the detailed chemical model [32] and a modified one-step Arrhenius-type model. It is seen that that the flame surface and the average flame velocity $S_f$ are larger in the case of a one-step model than it is for the detailed chemical model. However, the trend in the temporal variation of average flame velocities is the same for both chemical models. In particular, it is seen that the deceleration stage begins at almost the same time for both chemical models.

Figure 17 shows the calculated time evolution of the local velocities of the flame front on the axis and along $y = 0.16\,\text{cm}$ for the one-step model and at $y = 0.22\,\text{cm}$ for the detailed model. It is seen that the velocity of the flame front near the side wall for a one-step chemical model at $y = 0.16\,\text{cm}$, and for a detailed model at $y = 0.22\,\text{cm}$ exceeds the speed of the flame front at the center line at almost the same time in both cases also.

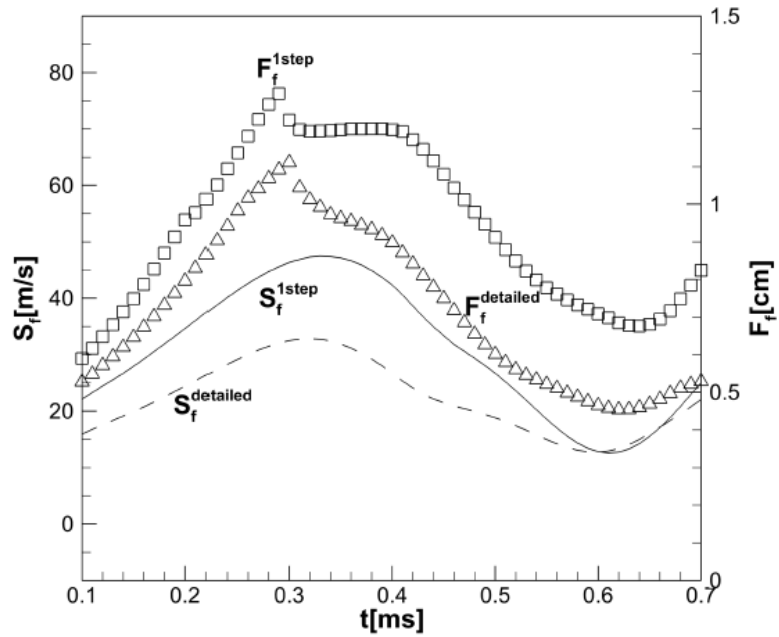

Figure 16. The time evolution of the flame surface area and the average flame velocity during the development of a tulip flame, calculated using a modified one-step model (squares) and a detailed chemical model [32] (dashed lines).

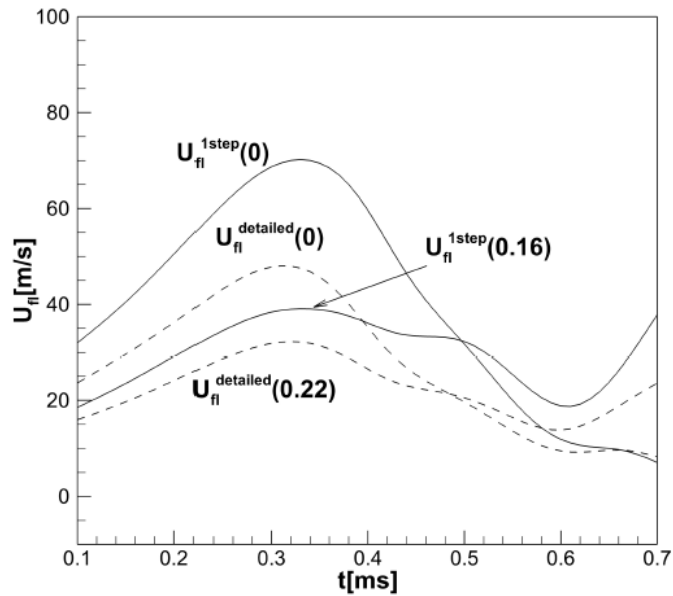

Figure 17. The time evolution of the hydrogen/air flame front velocities at the axis, $y = 0$ and near the sidewall, $y = 0.16\,cm$ and $y = 0.22\,cm$, during the development of the tulip flame. A modified one-step model (solid lines) and a detailed chemical model [32] (dashed lines).

The calculated schlieren images during the formation of a tulip flame obtained in simulations with the detailed chemical model, shown Figure 18 is quite similar to Fig. 4 obtained in simulations with the one-step chemical model, though there is a slight difference in the times of the development vortical structure behind the flame.

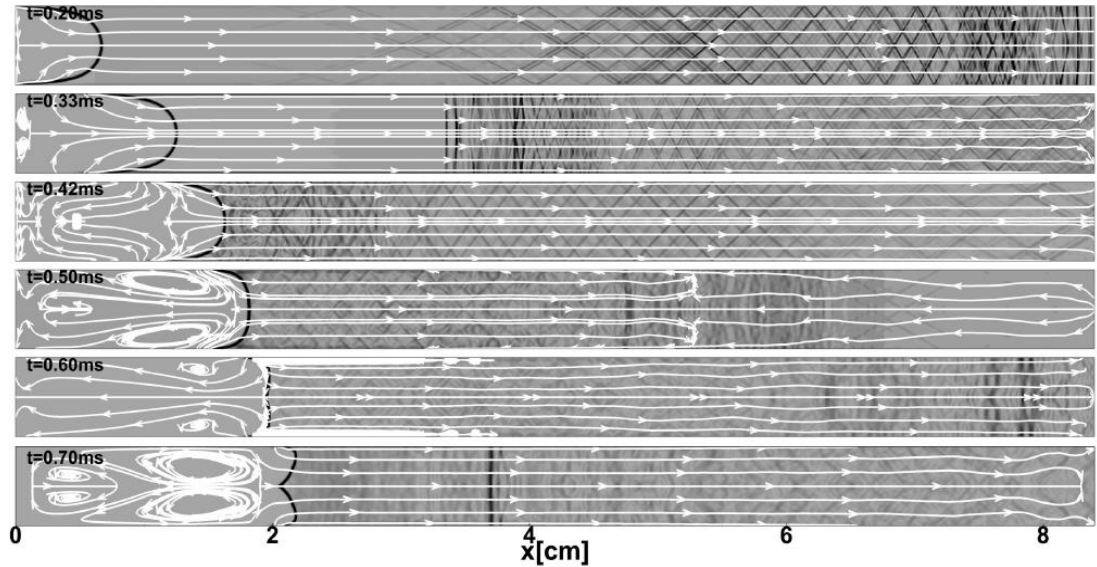

Figure 18. Sequences of schlieren images and streamlines in the flow ahead and behind the premixed hydrogen/air flame during the development a tulip flame calculated using a detailed chemical model.

The pressure profiles and rarefaction waves generated by the flame during the deceleration stage, as well as the flow velocity along the center line of the tube, are also very similar for simulations with a one-step or with a detailed chemical model, which can be considered is in favor of a purely gas dynamics mechanism of the formation of the tulip flame.

## 6. Summary and conclusions

This paper presents an analysis of the dynamics of the early stages of flame propagation in tubes with no-slip walls using numerical modeling to study the mechanism of tulip flame formation in closed and half-open tubes. The two-dimensional reactive Navier–Stokes equations coupled with either a one-step Arrhenius or detailed chemical models were solved using a high order numerical code and high-resolution simulations. The numerical simulations provided details of the flame dynamics and the processes through which the flame front inversion occurs from a convex shape with the cusp pointed in the unburned gas to a concave shape with a cusp pointing towards the burned gas. Simulations support the hypothesis proposed by Guenoche in 1964 that the tulip-shaped flame forms due to rarefaction waves generated by the flame after the flame underwent a sudden deceleration due to a decrease in the flame surface area caused by extinguishing of the rear part of the flame skirt on the sidewalls. Simulation also confirms the conclusion drawn from experimental studies of the formation of a tulip flame [1] that the formation of a tulip flame is a purely gas-dynamic process that is not associated with flame instabilities.Numerical simulations highlight the role of rarefaction waves generated by the flame during the deceleration phase, which are the main cause of the flame front inversion and the tulip flame formation. The mechanism of the tulip flame formation observed in simulations is an uneven decrease in the local velocities of the flame front, which is caused by a change in the velocity profile in the flow ahead of the flame by rarefaction waves created by the decelerating flame. The flow velocity ahead of the flame decreases more along the centerline of the tube, so that central part of the flame front begins to propagate more slowly than the side parts, which leads to inversion of the flame front. It should be noted that a decrease in the flow velocity ahead of the flame leads to an increase in the width of the boundary layer, which, in turn, determines the velocity profile in the flow ahead of the flame. As a result, the thickness of the formed tulip petals is approximately equal to the width of the boundary layer ahead of the flame. In the case of a fast flame in a tube with both ends closed, the rarefaction waves can cause a reversed flow of the unburned gas. The details of the flame front inversion during the tulip flame formation depend on the parameters of the problem: the laminar flame speed, the aspect ratio of a tube, and on the intensity of the ignition.

In general, the process of a tulip flame formation obtained in simulations with a one-step Arrhenius chemical model is similar to that obtained in simulations using a detailed chemical model. This can be seen as an indirect confirmation that the tulip flame formation is a purely gas dynamic process. However, since simulations with a one-step chemical model inevitably use also a simplified transport model, some details of the flow can be different. In particular, simulations of hydrogen/air flame with detailed chemical and transport models show the vortices created behind the flame near the sidewalls, but for the same conditions there is not vortices in simulations with a one-step chemical model. The vortices, which appear behind the flame near the sidewalls, are, apparently, the result of the interaction of flow in the direction along the centerline of the tube and the flow in the transverse direction. The vortices may expand with time or may disappear depending on the aspect ratio of the tube. The vortices in the flow behind the flame during the formation of the tulip flame appear rather as a result of purely gas-dynamic processes in the burned gas and are not associated with the mechanism of the tulip flame formation. However, it cannot be ruled out that vortices contribute to the creation of conditions for the inversion of the flame front, by creating a reverse flow on the centerline behind the flame, and therefore contribute to the formation of a tulip flame.

**Declaration of Competing Interest**

None.

**Acknowledgements**

This work was supported by National Natural Science Foundation of China [Grant Number 11732003 and U1830319].

M.L. acknowledges fruitful discussions with G. Sivashinsky and P. Clavin.

**Figure captions**

**Figure 1**. The time evolution of combustion wave velocity $S_f$ (solid line), local x-component of the flame front velocities $U_{fL}(x,0)$ at $y = 0$ and at $y = 0.16\,cm$, and the flame surface area (length) $F_f$ shown by squares. The tube width $D = 0.6\,cm$, $L/D = 14$, both ends closed.

**Figure 2.** Velocity profiles of the unburned gas immediately ahead of the flame at selected times.

**Figure 3.** Profiles of the x-component velocities of the flame front at selected times.

**Figure 4.** Sequences of calculated "schlieren images" and streamlines during the tulip flame formation for the condition of Figs. 1-3.

**Figure 5.** Pressure (solid line), temperature (dashed line) and flow velocity (dashed dotted lines) at the tube axis $y = 0$ during the tulip flame formation in hydrogen/air mixture.

**Figure 6.** The time evolution of the methane/air flame surface area $F_f$, the average flame speed $S_f$ (solid line), the local x-component of the flame front velocities $U_{fL}(0)$ at $y = 0$ and at $y = 0.16\,cm$. The tube with both ends closed, $D = 0.6\,cm$, $L/D = 14$.

**Figure 7.** The velocity profiles in the unburned gas immediately ahead of the methane/air flame at the selected times.

**Figure 8.** Local velocities of the methane/air flame front for the conditions in Fig. 7.

**Figure 9.** Sequences of schlieren images and streamlines during the tulip flame formation in a methane/air flame.

**Figure 10.** Pressure waves (solid line), temperature (dashed line) and flow velocity (dashed dotted lines) at the tube axis during the tulip flame formation in methane/air.

**Figure 11.** The evolution of the flame surface area $F_f$, the combustion wave velocity $S_f$ (solid line), and the local velocities of the flame front at $y = 0$ and $y = 0.2\,cm$.

**Figure 12.** x-component of the flow velocity profiles ahead of the hydrogen/air flame at selected times.

**Figure 13.** The evolution of x-component of the flame front velocity profiles at selected times.

**Figure 14.** Sequences of schlieren images and streamlines during the tulip flame formation in a hydrogen/air flame.

**Figure 15.** Profiles of pressure (solid line), temperature (dashed line) and flow velocity (dashed-dotted line) at the centerline of the half-open tube during the tulip shape formation of the hydrogen/air flame.

**Figure 16.** The time evolution of the flame surface area and the average flame velocity during the development of a tulip flame, calculated using a modified one-step model (squares) and a detailed chemical model [32] (dashed lines).

**Figure 17.** The time evolution of the hydrogen/air flame front velocities at the axis, $y = 0$ and near the sidewall, $y = 0.16\,cm$ and $y = 0.22\,cm$, during the development of the tulip flame. A modified one-step model (solid lines) and a detailed chemical model [32] (dashed lines).

**Figure 18.** Sequences of schlieren images and streamlines in the flow ahead and behind the premixed hydrogen/air flame during the development a tulip flame calculated using a detailed chemical model.